\documentclass[aps,prl,twocolumn,showpacs,preprintnumbers,superscriptaddress]{revtex4-2}
\usepackage{amsmath,amssymb,amsfonts}
\usepackage{graphicx}
\usepackage{amsthm}
\usepackage{dcolumn}
\usepackage{bm}
\usepackage{listings}
\usepackage{color}
\newcommand{\ket}[1]{|#1\rangle}

\newcommand{\comm}[2]{[\hat{#1},\hat{#2}]}
\newcommand{\Hhat}{\hat{H}}
\newcommand{\Chat}{\hat{C}}
\newcommand{\Khat}{\hat{K}}
\newcommand{\xhat}{\hat{x}}
\newcommand{\yhat}{\hat{y}}
\newcommand{\pxhat}{\hat{p}_x}
\newcommand{\pyhat}{\hat{p}_y}

\definecolor{seagreen}{rgb}{0.05, 0.65, 0.20}

\usepackage[colorlinks=true,linkcolor=blue,citecolor=blue,urlcolor=blue]{hyperref}
\begin{document}

\title{Ghost Degrees of Freedom Without Quantum Runaway: Exact Moment
Bounds from an Operator Conservation Law}

\author{Christopher Ewasiuk}
\affiliation{Department of Physics, University of California, Santa Cruz, Santa Cruz, CA 95064, USA}
\affiliation{Santa Cruz Institute for Particle Physics, University of California, Santa Cruz, Santa Cruz, CA 95064, USA}

\author{Stefano Profumo}
\affiliation{Department of Physics, University of California, Santa Cruz, Santa Cruz, CA 95064, USA}
\affiliation{Santa Cruz Institute for Particle Physics, University of California, Santa Cruz, Santa Cruz, CA 95064, USA}

\begin{abstract}
We prove an exact quantum conservation law for a harmonic
oscillator coupled to a ghost degree of freedom: a second
classical conserved quantity lifts to a quantum operator that
commutes with the Hamiltonian with no $\hbar$ corrections,
yielding a rigorous, state-independent upper bound on the mean squared phase-space radius for all time and every quantum state with finite initial second moments.
The proof uses only canonical commutation relations and the
Leibniz rule; it requires no confining potential, no spectral
assumptions, and no perturbative expansion.
The interaction studied here is bounded and vanishes at large
separations, the generic situation in effective field theory,
yet this suffices to guarantee quantum stability in the sense
of bounded second moments.
Three independent numerical frameworks (Heisenberg picture,
Schr\"{o}dinger picture, and Fock-space diagonalization)
confirm wavepacket confinement below the analytic bound,
a real energy spectrum, and Poisson level statistics numerically consistent
with an integrable structure.
The absence of a confining potential means the proof is silent
on spectral discreteness and the existence of a ground state;
those questions, addressed for polynomial confining interactions
in concurrent work~\cite{VacuumGhost2026,SpectralGhost2026},
remain open for the interaction class studied here and
represent the sharpest targets for future work.
Ghost quantum instability is therefore not an inevitable
consequence of a wrong-sign kinetic term but depends
critically on the interaction structure.
\end{abstract}

\maketitle

\section{Introduction}

Ghost fields, degrees of freedom with a wrong-sign kinetic term, appear across
a wide range of physically motivated theories: higher-derivative
gravity~\cite{Stelle1977}, Faddeev-Popov gauge fixing~\cite{FaddeevPopov1967},
dark-energy models~\cite{Carroll2003}, and the Pais-Uhlenbeck
oscillator~\cite{PaisUhlenbeck1950}.
The Ostrogradsky theorem~\cite{Ostrogradsky1850} and standard perturbation theory
both suggest that any ghost-containing theory is unstable: the ghost can radiate
energy into normal-sector modes without limit, and in quantum field theory the
vacuum decay rate naively diverges~\cite{ClineJeonMoore2004}.

Ghost fields arise concretely in the dark energy context.
Recent results from the Dark Energy Spectroscopic
Instrument (DESI) suggest a dark energy equation-of-state
parameter $w < -1$, or a dynamical $w(z)$ crossing the
phantom divide~\cite{DESI:2024mwx,DESI:2025zgx}, at a
statistical significance of roughly $2$--$3\sigma$ depending
on the dataset combination.
One class of scalar field models that accommodates $w < -1$
is the phantom field~\cite{Caldwell:1999ew}, which carries a
wrong-sign kinetic term; the standard quantum objection is
that such a ghost triggers catastrophic vacuum decay at a rate
that diverges in perturbation theory~\cite{ClineJeonMoore2004}. We note that this is not the only route: kinetic gravity braiding 
models~\cite{Deffayet:2010qz} and minimally modified 
gravity~\cite{MukohyamaMMG} achieve $w < -1$ without any ghost 
degree of freedom.
For the class of models where a ghost \emph{is} present, the
present Letter shows that the quantum instability objection
is not universal: catastrophic runaway depends on the
interaction structure, not merely on the sign of the kinetic
term.
We emphasize that our proof is quantum-mechanical rather
than field-theoretic, and the extension to phantom scalar
field theory remains a substantial open problem; nonetheless,
the result opens theoretical space for phantom dark energy
models that has hitherto been presumed closed.

Recent work by Deffayet, Mukohyama, and
Vikman~\cite{Deffayet2022,Deffayet2023} has challenged this
lore at the classical level.
Crucially, Refs.~\cite{Deffayet2022,Deffayet2023} prove not
merely Lyapunov stability of equilibria but the considerably
stronger result of global stability (Lagrange stability):
the motion is bounded in phase space for \emph{all} initial
conditions and for all time, distinguishing those works from
earlier studies that found only islands of stability.
Building on this classical foundation, we ask whether global
stability survives quantization.
Concurrently and independently, Deffayet, Fathe~Jalali,
Held, Mukohyama, and Vikman address the same quantization
question for a polynomial confining interaction, proving
existence of a unique vacuum, unitarity of time evolution,
and discreteness of the energy
spectrum~\cite{VacuumGhost2026,SpectralGhost2026}.
The present work studies a qualitatively different
interaction class: the potential $V_I$ of Eq.~\eqref{eq:VI}
is bounded everywhere and \emph{vanishes} at large
separations, so no confining walls enforce stability.
Our proof that second moments remain bounded in this setting
therefore requires no spectral input and no polynomial growth
at infinity; it holds for every quantum state by operator
algebra alone.

We answer affirmatively, with an important qualification: the stability we
establish is the boundedness of the mean squared phase-space radius, not spectral stability or the existence of a ground state, which are distinct and harder questions that we
discuss in the final section.  The key is an exact quantum conservation law: the second 
classical conserved quantity lifts to a quantum operator with 
no $\hbar$ corrections, yielding a rigorous, state-independent 
bound on the quantum dynamics. Self-adjointness of the Hamiltonian on $L^2(\mathbb{R}^2)$ is established below in 
Theorem~\hyperlink{prop:sa}{1} via the bounded perturbation theorem, which guarantees unitary evolution 
in the continuum and makes the bound~\eqref{eq:bound} rigorously applicable without additional assumptions. Numerical simulations in the Heisenberg picture, 
Schr\"{o}dinger picture, and Fock-space diagonalization 
consistently support the analytic result and reveal an 
integrable, ground-state-free spectrum.

\section{Model}

We study the two degree of freedom quantum system (natural units $\hbar = \omega = m = 1$)
\begin{equation}
  \Hhat = \tfrac{1}{2}\!\left(\pxhat^2 + \xhat^2\right)
         -\tfrac{1}{2}\!\left(\pyhat^2 + \yhat^2\right)
         + V_I(\xhat,\yhat),
  \label{eq:H}
\end{equation}
where $\xhat,\pxhat$ are the normal (positive-energy) oscillator operators,
$\yhat,\pyhat$ are the ghost operators (note the overall minus sign on the ghost
sector), and
\begin{equation}
  V_I(\xhat,\yhat) = \lambda\!\left[\!\left(\xhat^2-\yhat^2-1\right)^{\!2}+4\xhat^2\right]^{-1/2}.
  \label{eq:VI}
\end{equation}

The potential $V_I$ (the subscript denotes the interaction term) satisfies
$0 < |V_I| \le |\lambda|$ everywhere, since the expression under the square root
is bounded below by unity, as shown by the geometric identity~\eqref{eq:identity}
below. Canonical commutation relations hold within each sector; operators
from different sectors commute.  In the classical limit $\hbar\to 0$,
Eq.~\eqref{eq:H} reduces to the integrable system of Ref.~\cite{Deffayet2022},
whose Lyapunov stability was established in
Refs.~\cite{Deffayet2023}.

The Hamiltonian~\eqref{eq:H} requires care with respect to the 
standard $L^2$ inner product, because the ghost kinetic term 
flips the sign of the $\hat{p}^2_y$ contribution. As shown in 
Theorem \hyperlink{prop:sa}{1}, $\hat{H}$ is nevertheless self-adjoint on 
$\mathcal{D}(H_0)$ via the bounded perturbation theorem, since $V_I$ is a bounded 
self-adjoint multiplication operator. Stone's theorem then guarantees a strongly 
continuous unitary group $U(t) = e^{-it\hat{H}}$, making the bound~\eqref{eq:bound} rigorously 
applicable in the continuum. Throughout, we work with the truncated Fock space and 
discrete position grid as concrete computational frameworks.
\section{Exact quantum conservation law}
\label{sec:quant_con}
 
\subsection{From Classical to Quantum: The conserved operator}
 
The classical system~\cite{Deffayet2022} possesses, in addition to $H$, a second
conserved quantity
\begin{equation}
  C = K^2 + \left(p_x^2 + x^2\right) - \left(x^2-y^2-1\right)V_I(x,y),
  \label{eq:C_classical}
\end{equation}
where $K = p_y x + p_x y$ is the hyperbolic boost generator.  We promote this to
the quantum operator
\begin{equation}
  \Chat = \Khat^2 + \left(\pxhat^2 + \xhat^2\right) - \left(\xhat^2-\yhat^2-1\right)V_I(\xhat,\yhat),
  \label{eq:Chat}
\end{equation}
with $\Khat = \pyhat\xhat + \pxhat\yhat$.  Since $[\xhat,\pyhat]=[\yhat,\pxhat]=0$,
$\Khat$ is Hermitian.  There is no operator-ordering ambiguity in $\Chat$: both
$V_I$ and $(\xhat^2-\yhat^2-1)$ are functions of commuting position operators.
 
The key geometric identity
\begin{equation}
  (x^2 - y^2 - 1)^2 + 4x^2 - (y^2 - x^2)^2 = 2x^2 + 2y^2 + 1 \geq 1,
  \label{eq:identity}
\end{equation}
which holds for all $(x,y)\in\mathbb{R}^2$, yields the pointwise bound
\begin{equation}
  \left|(y^2-x^2)\right| V_I(x,y) < |\lambda|
  \quad \forall\,(x,y)\in\mathbb{R}^2,
  \label{eq:pointwise}
\end{equation}
which promotes to the operator inequality
$-|\lambda|\hat{1} < (\hat{y}^2-\hat{x}^2)V_I(\hat{x},\hat{y}) < |\lambda|\hat{1}$.
We show in the next section that $[\Chat,\Hhat]=0$ exactly for all
$\hbar>0$, with no operator-ordering remainder, and that this conservation law
yields the main result for all $t_b > 0$:
\begin{equation}
\bigl\langle \hat{x}^2 + \hat{y}^2 + \hat{p}_x^2 + \hat{p}_y^2
\bigr\rangle(t_b)
\;\leq\;
\bigl\langle \hat{x}^2 + \hat{y}^2 + \hat{p}_x^2 + \hat{p}_y^2
\bigr\rangle(0)
\;+\; 4|\lambda|.
\label{eq:bound}
\end{equation}
This is a statement about second moments of the quantum state, not its
spatial support: the wavefunction can in principle develop tails outside any
compact region while $\langle\hat{x}^2+\hat{y}^2\rangle$ remains bounded.
Subject to this qualification, the bound holds for every quantum state, pure
or mixed, coherent or Fock; independently of $\hbar$, the wavefunction, and
the energy.  It reduces to the classical bound of Ref.~\cite{Deffayet2022}
as $\hbar\to 0$, since no $\hbar$-dependent terms appear in Eq.~\eqref{eq:bound}.
 
\section{Operator-theoretic formulation and conserved moment bound}
\label{sec:optheory}
 
We place the model on the Hilbert space $\mathcal{H} = L^2(\mathbb{R}^2)$,
with canonical operators
\[
x,\, y \ \text{(multiplication)}, \qquad p_x = -i\partial_x,\quad p_y = -i\partial_y,
\]
defined on the usual Sobolev domains. Throughout we set $\hbar = 1$.
Define the free operator
\begin{equation}
H_0 := \frac{1}{2}(p_x^2 + x^2) - \frac{1}{2}(p_y^2 + y^2),
\end{equation}
the interaction $V_I(x,y) = \lambda \big[(x^2 - y^2 - 1)^2 + 4x^2\big]^{-1/2}$,
the boost operator $K := x p_y + y p_x$, the conserved operator
\begin{equation}
C := K^2 + (p_x^2 + x^2) - (x^2 - y^2 - 1)V_I(x,y),
\end{equation}
the positive semi-definite operator
\begin{equation}
\Sigma := K^2 + \tfrac{1}{2}(p_x^2 + x^2) + \tfrac{1}{2}(p_y^2 + y^2) \geq 0,
\end{equation}
and the conserved combination $E := C - H$.  The finite Hermite span
$\mathcal{D}_{\mathrm{fin}} := \mathrm{span}\{h_n(x)h_m(y): n,m\geq 0\}$
is invariant under $x, y, p_x, p_y$, and hence under $H$, $K$, and $C$.
 
\medskip

 \begin{figure}[t]
  \centering
  \includegraphics[width=\columnwidth]{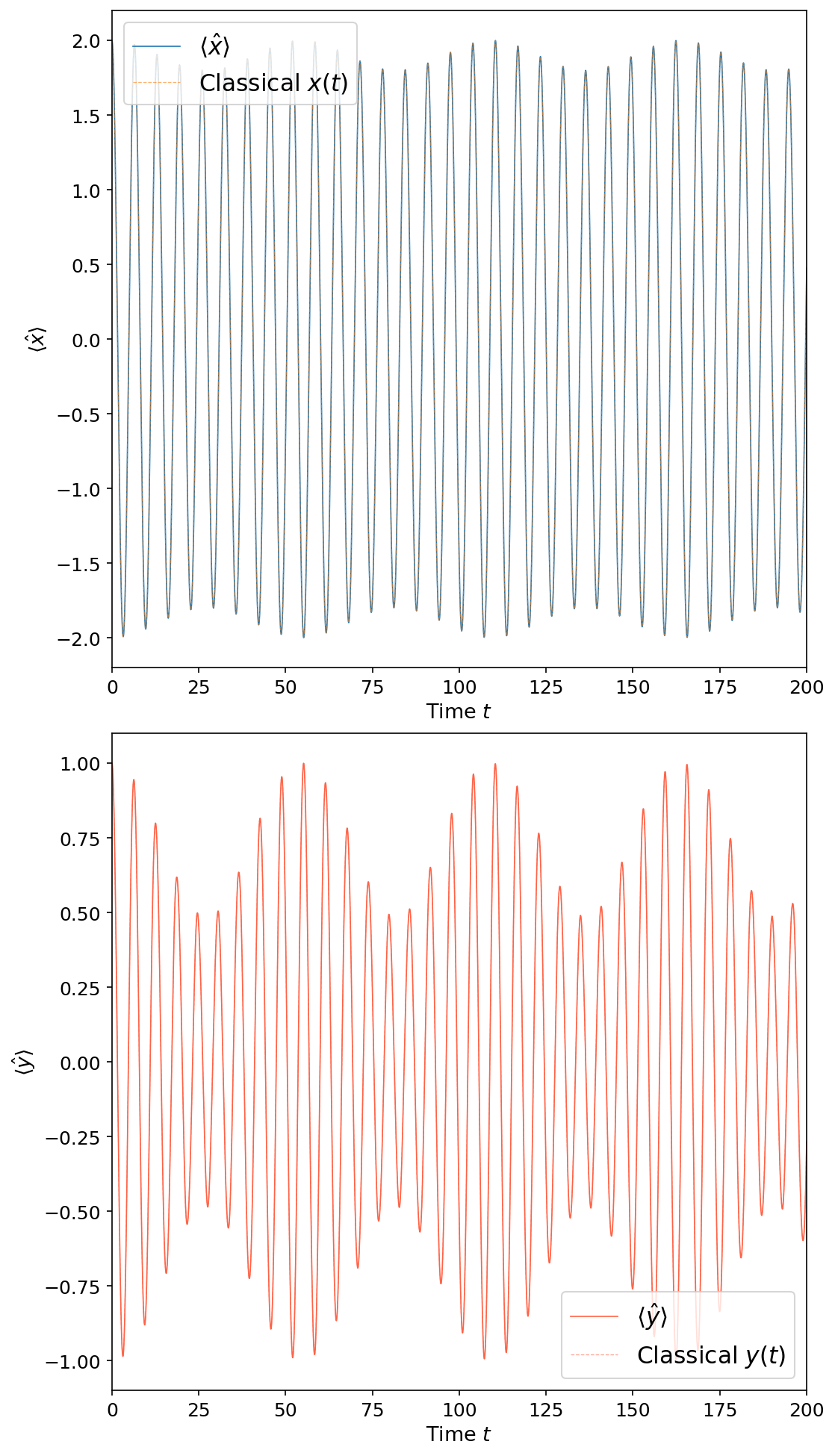}
  \caption{Heisenberg-picture Ehrenfest moments for $\lambda=1/3$, RK4,
    $\Delta t=0.02$, t = 500 (t cropped to $\in [0,200]$). Mean positions $\langle\hat{x}\rangle(t)$
    (left) and $\langle\hat{y}\rangle(t)$ (right) overlaid on the classical
    orbits (dashed); the curves are indistinguishable, confirming that
    classical stability survives at leading order in $\hbar$.}
  \label{fig:heisenberg}
\end{figure}

\noindent

\hypertarget{prop:sa}{}
\textbf{Theorem 1 (Self-adjointness).}
The Hamiltonian
$H := H_0 + V_I$
is self-adjoint on $D(H_0)$ and essentially self-adjoint on
$\mathcal{D}_{\mathrm{fin}}$.
 
\noindent
\emph{Proof.}
Let $A := \frac{1}{2}(-\partial_t^2 + t^2)$ denote the one-dimensional harmonic
oscillator on $L^2(\mathbb{R})$, with domain $D(A)$. Then
$H_0 = A \otimes I - I \otimes A$
on $\mathcal{H} \cong L^2(\mathbb{R})_x \otimes L^2(\mathbb{R})_y$.
Since $A\otimes I$ and $I\otimes A$ strongly commute, $H_0$ is self-adjoint on
$D(H_0) = D(A\otimes I)\cap D(I\otimes A)$.
The identity $(x^2-y^2-1)^2+4x^2\geq 1$ implies $|V_I(x,y)|\leq|\lambda|$
for all $(x,y)\in\mathbb{R}^2$, so $V_I$ is a bounded self-adjoint multiplication
operator. Hence $H$ is self-adjoint on $D(H_0)$ by the bounded perturbation
theorem. Essential self-adjointness on $\mathcal{D}_{\mathrm{fin}}$ follows
because $\mathcal{D}_{\mathrm{fin}}$ is a core for $H_0$ and is invariant
under $V_I$.\qed
 
\medskip
 
By Stone's theorem, $H$ generates a strongly continuous unitary group
$U(t) = e^{-itH}$, making the bound~\eqref{eq:bound} rigorously applicable
in the continuum.
 
\medskip
 
\noindent
\textbf{Theorem 2 (Conserved moment bound).}
Let $\psi_0 \in D(\Sigma^{1/2})$, and define $\psi_t := e^{-itH}\psi_0$.
Then for all $t\in\mathbb{R}$,
\begin{equation}
\langle \psi_t, \Sigma\, \psi_t \rangle
\leq
\langle \psi_0, \Sigma\, \psi_0 \rangle + 2|\lambda|.
\label{eq:sigbound}
\end{equation}
Consequently, Eq.~\eqref{eq:bound} holds.
 
\noindent
\emph{Proof.}
On $\mathcal{D}_{\mathrm{fin}}$, all operators are well-defined and all
commutators reduce using the canonical relations
$[x,p_x]=i$, $[y,p_y]=i$, with all cross-commutators vanishing, together with
$[p_x,f]=-i\,\partial_x f$ and $[p_y,f]=-i\,\partial_y f$
for smooth multiplication operators $f(x,y)$.
Since $C$ is at most quadratic in momenta, every term in $[C,H]$ involves at
most first derivatives of $V_I$ via the Leibniz rule, and the correspondence
$[\hat{A},\hat{B}]=i\hbar\{A,B\}_{\mathrm{PB}}$ is exact with no higher-order
remainder. A direct computation then gives $[C,H]=0$ by exact cancellation,
corresponding to the classical identity $\{C,H\}_{\mathrm{PB}}=0$
(Ref.~\cite{Deffayet2022}); the explicit operator-ordered derivation,
including the step-by-step cancellation enforced by the geometric
identity~\eqref{eq:identity}, is given in the Supplemental
Material~\cite{SM}.
Standard approximation arguments extend this to all of $D(H_0)$, giving
$\langle E\rangle_t = \langle E\rangle_0$ under unitary evolution.
 
A direct algebraic computation yields
$E = \Sigma + (y^2-x^2)V_I(x,y)$.
Since $(y^2-x^2)V_I$ is a bounded self-adjoint operator with norm $\leq|\lambda|$
by Eq.~\eqref{eq:pointwise}, we have the quadratic-form inequalities
\begin{equation}
\Sigma \leq E + |\lambda|\,I, \qquad \Sigma \geq E - |\lambda|\,I.
\end{equation}
Applying these at times $0$ and $t$,
\[
\langle\Sigma\rangle_t
\leq \langle E\rangle_t + |\lambda|
= \langle E\rangle_0 + |\lambda|
\leq \langle\Sigma\rangle_0 + 2|\lambda|.
\]
Finally, since $\hat{K}^2\geq 0$ implies
$x^2+y^2+p_x^2+p_y^2 \leq 2\Sigma$ as quadratic forms,
setting $t_a=0$ gives Eq.~\eqref{eq:bound}. \qed

Consequently,
$\tfrac{d}{dt}\langle\hat{C}\rangle = \tfrac{1}{i\hbar}
\langle[\hat{C},\hat{H}]\rangle = 0$
for every quantum state, provided the time evolution is unitary;
we return to this point in the \hyperlink{sec:discussion}{Discussion}.

\section{Numerical verification}

As a semiclassical consistency check, we integrate the Heisenberg equations
of motion $i\hbar\,d\hat{q}/dt = [\Hhat,\hat{q}]$,
\begin{align}
  \dot{\xhat} &= \pxhat, \label{eq:EOM1}\\
  \dot{\pxhat} &= -\xhat + \lambda\,\frac{2\xhat^3-2\xhat\yhat^2+2\xhat}
                  {\left[(\xhat^2-\yhat^2-1)^2+4\xhat^2\right]^{3/2}},
                  \label{eq:EOM2}\\
  \dot{\yhat} &= -\pyhat, \label{eq:EOM3}\\
  \dot{\pyhat} &= \phantom{-}\yhat - \lambda\,\frac{2\yhat\xhat^2-2\yhat^3-2\yhat}
                  {\left[(\xhat^2-\yhat^2-1)^2+4\xhat^2\right]^{3/2}}.
                  \label{eq:EOM4}
\end{align}
The reversed sign $\dot{\hat{y}}=-\hat{p}_y$ (versus $\dot{\hat{x}}=+\hat{p}_x$)
is the operator-level signature of the ghost. At the Ehrenfest level,
$\langle\hat{x}\rangle(t)$ and $\langle\hat{y}\rangle(t)$ track the classical
orbits of Refs.~\cite{Deffayet2022,Deffayet2023} to within numerical precision
throughout $t\in[0,500]$ (Fig.~\ref{fig:heisenberg}), confirming that the
classical stability survives at leading order in $\hbar$.
\begin{figure}[t]
  \centering
  \includegraphics[width=\columnwidth]{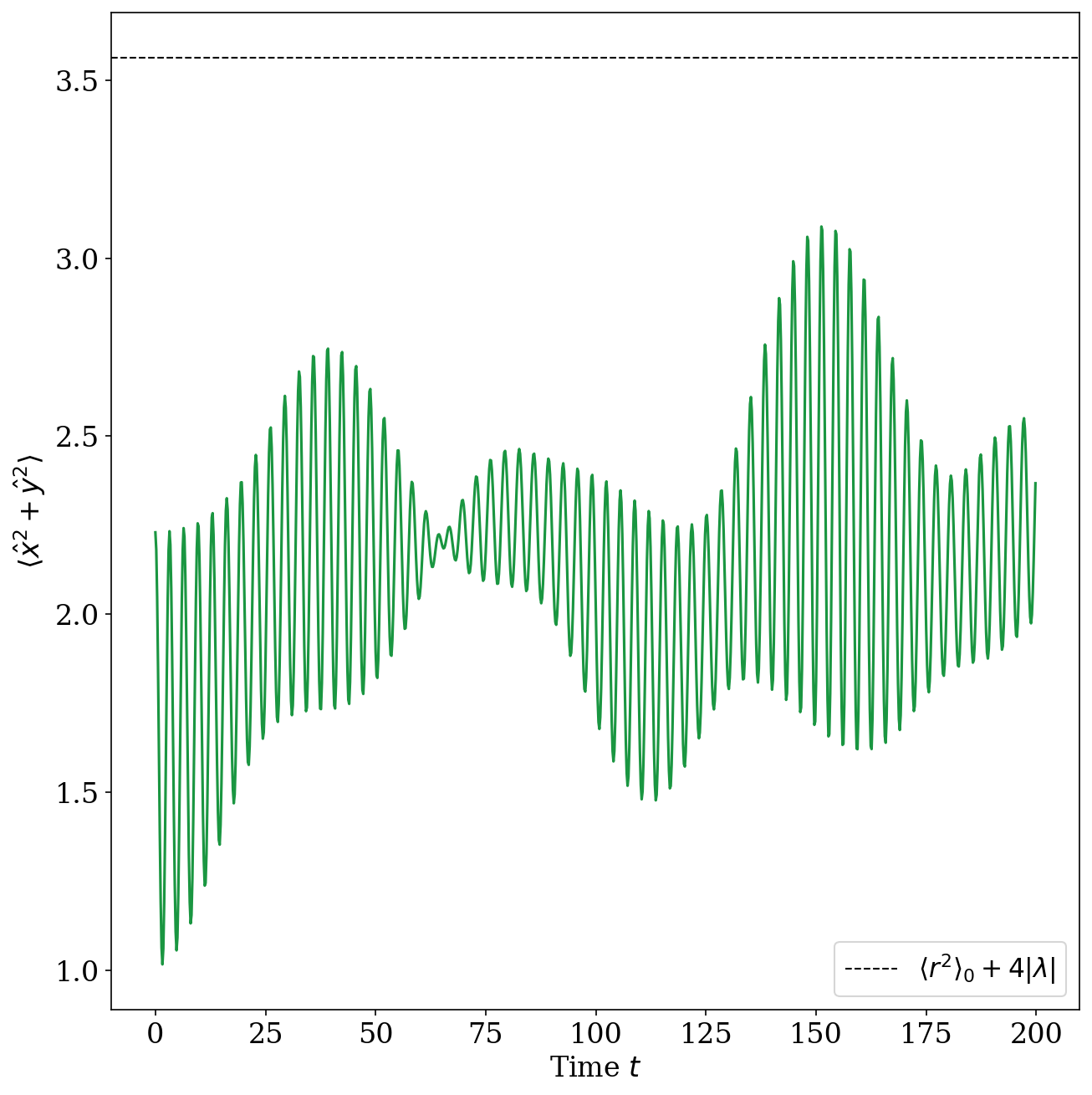}
  \caption{Schr\"{o}dinger-picture confirmation of the moment bound~\eqref{eq:bound}
    for $\lambda=1/3$, $N=128^2$, $\Delta t=5\times10^{-3}$, $T=200$,
    starting from a Gaussian wavepacket centered at
    $(x_0,y_0)=(1.0,0.5)$ with width $\sigma=0.7$.
    The mean squared radius $\langle\hat{x}^2+\hat{y}^2\rangle(t)$
    oscillates with no secular growth, remaining well below the analytic
    ceiling $\langle r^2\rangle_0+4|\lambda|$ (dashed) throughout the run.}
  \label{fig:schrodinger}
\end{figure}

\begin{figure}[t]
  \includegraphics[width=\columnwidth]{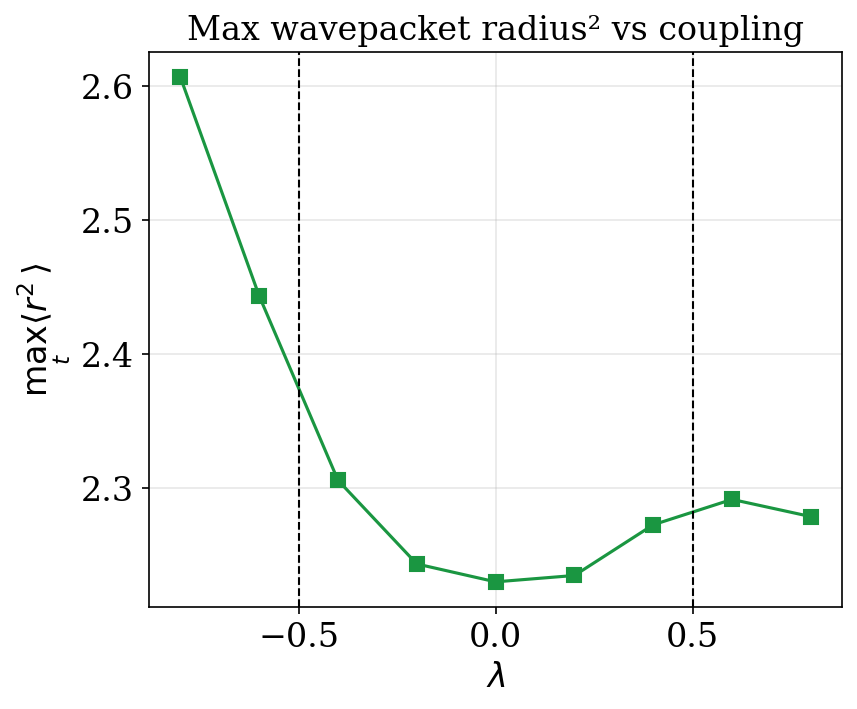}
  \caption{Maximum mean squared radius $\max_t\langle\hat{x}^2+\hat{y}^2\rangle$
  over $t\in[0,200]$ as a function of coupling strength $\lambda$, obtained
  from Schr\"{o}dinger-picture propagation on a $128\times128$ grid.
  Wavepacket confinement persists across the full range
  $\lambda\in[-0.8,+0.8]$ with no runaway and no sharp transition at the
  classical Lyapunov boundary $|\lambda|=1/2$ (dashed lines).
  The maximum remains well below the analytic ceiling
  $\langle r^2\rangle(0)+4|\lambda|$ of Eq.~\eqref{eq:bound} throughout,
  confirming that the bound is satisfied but not saturated at these coupling
  strengths.}
  \label{fig:lambda_scan}
\end{figure}

We also propagate the time-dependent Schr\"{o}dinger equation directly, via
the Crank--Nicolson scheme~\cite{CrankNicolson1947}, which preserves
discrete unitarity so that any residual norm drift is double-precision
rounding rather than a physical instability. The Gaussian wavepacket of
Fig.~\ref{fig:schrodinger} exhibits $\langle\hat{x}^2+\hat{y}^2\rangle(t)$
bounded in $[1.02, 2.61]$ with no secular growth, and the coupling scan of
Fig.~\ref{fig:lambda_scan} extends this confinement across
$\lambda\in[-0.8,+0.8]$ including well past the classical Lyapunov boundary
$|\lambda|=1/2$.

\section{Energy spectrum}

\begin{figure*}[htbp]
    \centering
    \includegraphics[width=\textwidth]{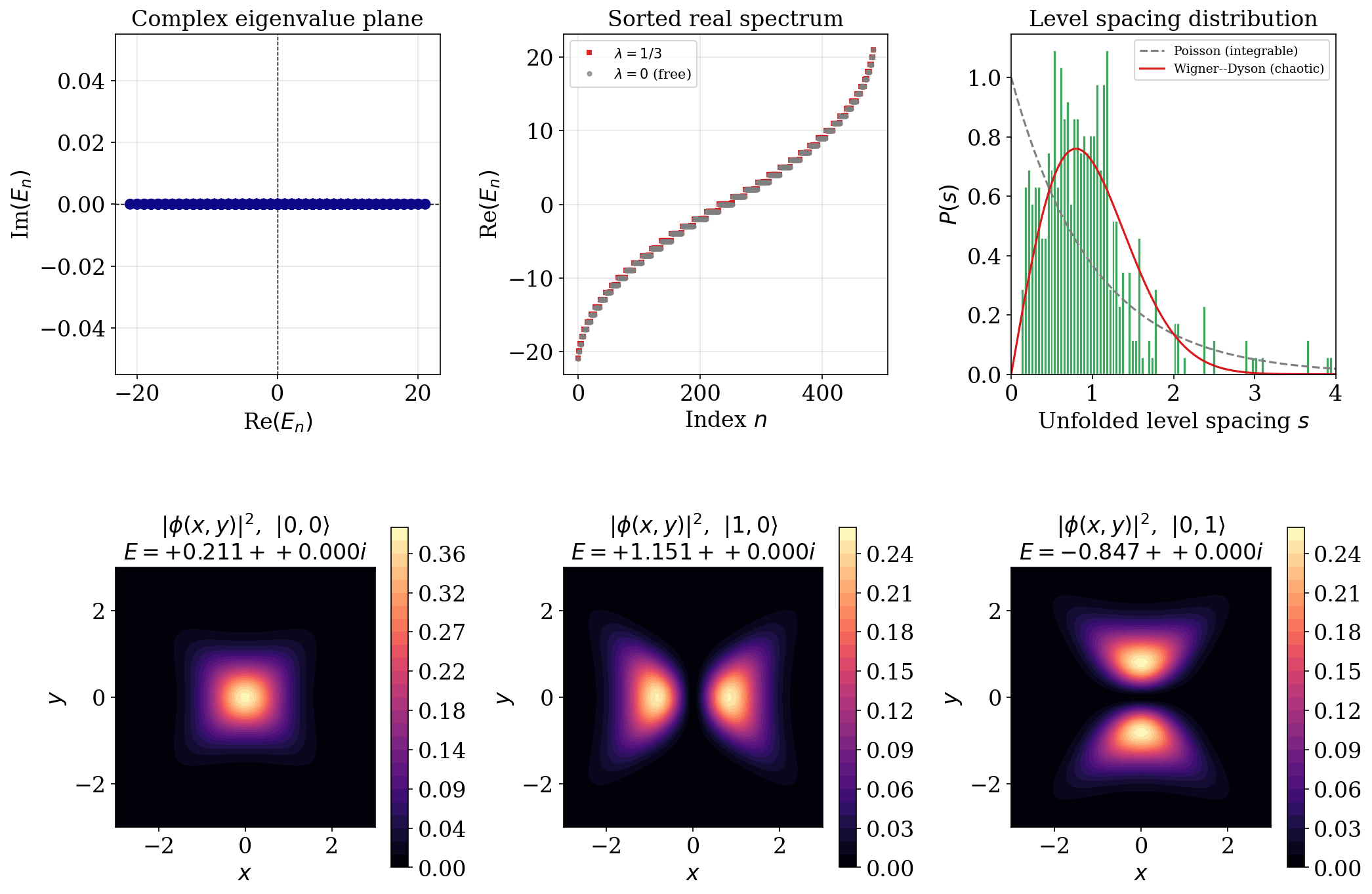}

    \caption{Energy spectrum of the ghost-coupled oscillator for
$\lambda=1/3$, Fock basis with $N_{\rm max}=21$ (484 states).
\emph{Top left}: all 484 eigenvalues in the complex plane;
all lie on the real axis within LAPACK precision.
\emph{Top center}: sorted real eigenvalues for $\lambda=0$
(grey) and $\lambda=1/3$ (red); shifts are $\lesssim 10^{-4}$,
consistent with $V_I$ acting perturbatively.
\emph{Top right}: unfolded level-spacing distribution of the
intra-multiplet splittings introduced by $V_I$, compared with
Poisson $P(s)=e^{-s}$ (dashed) and Wigner-Dyson
$P(s)\propto se^{-\pi s^2/4}$ (solid red).
Within each degenerate multiplet the distribution is consistent
with Wigner-Dyson, as expected from degenerate perturbation
theory; the global inter-multiplet structure remains Poisson,
consistent with an integrable system.
\emph{Bottom row}: probability density $|\phi_n(x,y)|^2$ for
three low-lying eigenstates identified by maximum overlap with
$|0,0\rangle$, $|1,0\rangle$, and $|0,1\rangle$, showing the
expected nodal structure of harmonic oscillator product states.}
    \label{fig:spectrum}
\end{figure*}

At $\lambda=0$ the Hamiltonian separates and the spectrum is
$E^{(0)}_{n_x n_y} = \hbar\omega(n_x - n_y)$, unbounded in both directions
with no ground state --- a structural feature of the ghost sector that the
interaction does not cure. For $\lambda\neq 0$ we diagonalize $\Hhat$ in the
truncated Fock basis $\{\ket{n_x}\otimes\ket{n_y}:n_x,n_y\le N_{\rm max}=21\}$
(484 states). The integrable structure implied by $\hat{C}$ predicts Poisson
inter-multiplet level spacings, and this is what we find: the integer-spaced
ladder of degenerate subspaces survives the interaction intact, consistent
with the Poisson statistics expected for an integrable system
(Fig.~\ref{fig:spectrum}).

All eigenvalues lie on the real axis within LAPACK precision, consistent
with the PT-symmetry analysis of the continuum
operator~\cite{BenderBoettcher1998}; $V_I$ produces shifts $\lesssim 10^{-4}$
and opens no gap. (The minimum observed gap $\sim 10^{-11}$ is numerical
round-off in near-degenerate diagonalization, five orders of magnitude below
the perturbative splitting scale.) Low-lying eigenstates are spatially
localized within the central region of the grid, providing a spectral
counterpart to the dynamical wavepacket confinement of Eq.~\eqref{eq:bound}.
The bounded phase-space moments established above therefore genuinely
coexist with an unbounded, ground-state-free spectrum.


\hypertarget{sec:discussion}{}
\section{Discussion and conclusions}


The central result of this Letter is analytic and exact: the operator
$\hat{C}$ defined in Eq.~\eqref{eq:Chat} satisfies
$[\hat{C}, \hat{H}] = 0$ with no $\hbar$ corrections, as a direct
consequence of canonical commutation relations and the Leibniz rule alone.
From this single fact, elementary operator algebra yields
the bound~\eqref{eq:bound}: the mean squared phase-space radius
$\langle \hat{x}^2 + \hat{y}^2 + \hat{p}_x^2 + \hat{p}_y^2 \rangle$ is bounded
for all time, for every quantum state with finite initial second moments,
at all coupling strengths, provided the time evolution is unitary.

We now state precisely what the bound does and does not establish.

\textit{What is proven.} The bound~\eqref{eq:bound} is a rigorous statement
about second moments of the quantum state.  It holds for every pure or mixed
state, coherent or Fock, and reduces to the classical bound of
Ref.~\cite{Deffayet2022} in the limit $\hbar \to 0$, since no $\hbar$-dependent
terms appear in Eq.~\eqref{eq:bound}.  There are no operator-ordering
ambiguities in the proof and no perturbative remainder: the commutator
$[\hat{C}, \hat{H}] = 0$ is exact for all $\hbar > 0$.

\textit{What is not proven.} The bound does not imply 
spectral stability, the existence of a ground state, or 
spatial localization of the wavefunction beyond the control 
of second moments: in principle the wavefunction can develop 
tails outside any compact region while $\langle \hat{x}^2 + 
\hat{y}^2 \rangle$ remains bounded. The question of whether 
$\hat{H}$ admits a self-adjoint extension on $L^2(\mathbb{R}^2)$ 
is resolved affirmatively in Theorem \hyperlink{prop:sa}{1}, which 
establishes that unitary evolution holds in the continuum and 
that the bound \eqref{eq:bound} applies literally. What remains open is the construction of the positive-definite $\mathcal{CPT}$ inner product from the
eigenstates of $\hat{H}$, and the extension of these results
to ghost-coupled field theories including Lee-Wick
models~\cite{LeeWick1969,LeeWick1970} and the
Pais-Uhlenbeck oscillator~\cite{PaisUhlenbeck1950}.
Concurrent work~\cite{VacuumGhost2026,SpectralGhost2026}
resolves two of the questions left open here, establishing
a unique vacuum and a discrete (non-dense) spectrum for
ghost systems with polynomial confining interactions.
Whether spectral discreteness accompanies bounded second
moments for the bounded, vanishing interaction $V_I$ of
Eq.~\eqref{eq:VI} is the sharpest open question left by
the present work: the operator $\hat{C}$ predicts Poisson
inter-multiplet statistics (confirmed numerically in
Fig.~\ref{fig:spectrum}), but a proof that the spectrum is not dense for
non-confining interactions remains out of reach with
current methods.

\textit{Relation to the Ostrogradsky theorem.}  The most natural objection to
the preceding is that the Ostrogradsky theorem guarantees the Hamiltonian is
unbounded below, and our spectrum confirms this.  The theorem does not,
however, preclude bounded phase-space moments, and our result shows
explicitly that the two can coexist.  Ostrogradsky instability and bounded
second moments are statements about different properties of the dynamics,
and the interaction structure of $V_I$ is precisely what allows them to be
simultaneously satisfied.

\textit{Structural relevance to Lee-Wick and Pais-Uhlenbeck.}  The boundedness
of $V_I$ and the existence of $\hat{C}$ both depend on the specific
hyperbolic-boost-covariant combination $(x^2-y^2-1)^2+4x^2$, not on the
wrong-sign kinetic term alone.  Lee-Wick models~\cite{LeeWick1969,LeeWick1970}
and the Pais-Uhlenbeck oscillator~\cite{PaisUhlenbeck1950} acquire their
ghost sector from higher-derivative kinetic terms that, after reduction to a
two-oscillator form, do not generically share this symmetry structure; whether physically motivated interactions in those theories
can be arranged to preserve an analogous conserved operator
is an open question that the present construction makes sharp.
A natural intermediate step toward the field-theoretic
extension is the classical field theory program of
Refs.~\cite{HeldDeffayet2025,Held2025}, which establishes
small-data global stability for ghostly field configurations
in $(N+1)$ dimensions: the dynamical decoupling mechanism
responsible for classical stability in the mechanical model
persists in the field-theoretic setting, lending support to
the conjecture that an analogous quantum operator bound
may survive the passage to infinitely many degrees of freedom.
The key obstacles specific to field theory, namely
renormalization, the absence of a conserved $\hat{C}$ in
the continuum limit, and the divergence of the naive
vacuum decay rate, remain open and represent the sharpest
targets for future work.

\textit{Relation to prior benign-ghost and pseudo-Hermiticity approaches.}
Two bodies of work have previously argued that ghost-coupled systems can be
well-behaved. Smilga and collaborators~\cite{Smilga2005,DamourSmilga2022}
identified classes of ``benign ghost'' mechanical
systems whose classical motion remains bounded and whose quantum evolution
is unitary, typically by exhibiting specific interaction forms or by
restricting to islands of stability in phase space. Mostafazadeh and
others in the pseudo-Hermitian / $\mathcal{PT}$-symmetric quantum mechanics
program~\cite{Mostafazadeh2010} obtain a stable,
unitary Pais-Uhlenbeck quantization by redefining the Hilbert-space inner
product, replacing the standard $L^2$ metric with a positive-definite
$\mathcal{CPT}$-like inner product with respect to which the Hamiltonian
is Hermitian. Our result is logically distinct from both approaches.
The Hamiltonian studied here is self-adjoint on the standard $L^2(\mathbb{R}^2)$
inner product (Theorem~\hyperlink{prop:sa}{1}), so unitarity is inherited
from the bounded-perturbation theorem rather than from a redefinition of
the inner product; the bound~\eqref{eq:bound} then follows from an exact
operator identity $[\hat{C},\hat{H}]=0$ rather than from exact solvability
or confinement to a stability island. To our knowledge, this is the first
rigorous, state-independent, non-perturbative upper bound on the mean
squared phase-space radius for a ghost-coupled quantum system that is not
exactly solvable, obtained without modifying the inner product.

\textit{Role of the numerics.} With continuum unitarity established
analytically (Theorem~\hyperlink{prop:sa}{1}), the numerical results play
a complementary rather than a logically necessary role: the Heisenberg,
Schr\"{o}dinger, and Fock-space calculations collectively confirm that
the analytic bound~\eqref{eq:bound} and the integrable structure implied
by $\hat{C}$ are reflected in the actual dynamics --- at the Ehrenfest,
wavepacket, and spectral levels, respectively.

Together these results sharpen the message of
Refs.~\cite{Deffayet2022,Deffayet2023}: ghost instability is not an inescapable
consequence of a wrong-sign kinetic term, but depends critically on the
interaction structure. The present work provides the first rigorous
quantum-mechanical demonstration of this fact, extending the classical
Lyapunov result to an exact operator bound with no $\hbar$ corrections and no
perturbative remainder.

Open questions for future work include: whether
$\mathcal{CPT}$ norm conservation holds pointwise in the
continuum theory; a grid-refinement study of the
Schr\"{o}dinger propagation to verify convergence of
$\langle R^2\rangle(t)$ as $\Delta x \to 0$; whether
spectral discreteness can be established for bounded,
vanishing interactions of the form studied here; and the
extension of these results to ghost-coupled field theory,
including Lee-Wick models and the Pais-Uhlenbeck
oscillator~\cite{PaisUhlenbeck1950}.
The companion program of
Refs.~\cite{VacuumGhost2026,SpectralGhost2026} shows that
confining polynomial interactions yield discrete spectra,
raising the question of whether the dynamical decoupling
mechanism responsible for our moment bound is also
sufficient to discretize the spectrum without confining
walls.

\textit{Note added.}
While finalizing this work we became aware of two companion
papers by Deffayet, Fathe~Jalali, Held, Mukohyama, and
Vikman submitted
concurrently~\cite{VacuumGhost2026,SpectralGhost2026}.
Reference~\cite{VacuumGhost2026} canonically quantizes a
globally stable ghost oscillator with a polynomial confining
interaction, proves unitarity, constructs a unique positive
ground state, and establishes bounded second moments via a
positive-definite integral of motion $\hat{H}_\uparrow$.
Reference~\cite{SpectralGhost2026} proves via separability
theory and WKB analysis that the energy spectrum need not
be continuous or dense, constructing examples with exactly
one finite accumulation point or none.
The interaction studied in those works is polynomial and
confining; ours is bounded and vanishes at large separations.
Our bound is state-independent and requires no spectral
input; the two approaches are therefore complementary rather
than duplicative.
Reference~\cite{VacuumGhost2026} notes our work in its
closing remark.

\begin{acknowledgments}
SP is partly supported by the U.S.\ Department of Energy grant number DE-SC0010107.
\end{acknowledgments}

\bibliographystyle{apsrev4-2}
\bibliography{bibliography}

@article{Deffayet2022,
  author  = {Deffayet, C{\'e}dric and Mukohyama, Shinji and Vikman, Alexander},
  title   = {Ghosts Without Runaway Instabilities},
  journal = {Phys.\ Rev.\ Lett.},
  volume  = {128},
  pages   = {041301},
  year    = {2022},
  doi     = {10.1103/PhysRevLett.128.041301},
  eprint  = {2108.06294},
  archivePrefix = {arXiv},
  primaryClass  = {gr-qc}
}

@article{Deffayet2023,
  author  = {Deffayet, C{\'e}dric and Held, Aaron and Mukohyama, Shinji and Vikman, Alexander},
  title   = {Global and Local Stability for Ghosts Coupled to Positive Energy
             Degrees of Freedom},
  journal = {J.\ Cosmol.\ Astropart.\ Phys.},
  volume  = {2023},
  pages   = {031},
  year    = {2023},
  doi     = {10.1088/1475-7516/2023/11/031},
  eprint  = {2305.09631},
  archivePrefix = {arXiv},
  primaryClass  = {gr-qc}
}

@article{Stelle1977,
  author  = {Stelle, K. S.},
  title   = {Renormalization of Higher-Derivative Quantum Gravity},
  journal = {Phys.\ Rev.\ D},
  volume  = {16},
  pages   = {953},
  year    = {1977},
  doi     = {10.1103/PhysRevD.16.953}
}

@article{FaddeevPopov1967,
  author  = {Faddeev, L. D. and Popov, V. N.},
  title   = {Feynman Diagrams for the {Yang-Mills} Field},
  journal = {Phys.\ Lett.\ B},
  volume  = {25},
  pages   = {29},
  year    = {1967},
  doi     = {10.1016/0370-2693(67)90067-6}
}

@article{Carroll2003,
  author  = {Carroll, Sean M. and Hoffman, Mark and Trodden, Mark},
  title   = {Can the Dark Energy Equation-of-State Parameter $w$ Be Less
             Than $-1$?},
  journal = {Phys.\ Rev.\ D},
  volume  = {68},
  pages   = {023509},
  year    = {2003},
  doi     = {10.1103/PhysRevD.68.023509},
  eprint  = {astro-ph/0301273},
  archivePrefix = {arXiv}
}

@article{PaisUhlenbeck1950,
  author  = {Pais, A. and Uhlenbeck, G. E.},
  title   = {On Field Theories with Non-Localized Action},
  journal = {Phys.\ Rev.},
  volume  = {79},
  pages   = {145},
  year    = {1950},
  doi     = {10.1103/PhysRev.79.145}
}

@article{Ostrogradsky1850,
  author  = {Ostrogradsky, M. V.},
  title   = {M{\'e}moires sur les {\'e}quations diff{\'e}rentielles relatives
             au probl{\`e}me des isop{\'e}rim{\`e}tres},
  journal = {M{\'e}m.\ Acad.\ Imp.\ Sci.\ St.-P{\'e}tersbourg},
  volume  = {6},
  pages   = {385},
  year    = {1850}
}

@article{ClineJeonMoore2004,
  author  = {Cline, James M. and Jeon, Sangyong and Moore, Guy D.},
  title   = {The Phantom Menaced: Constraints on Low-Energy Effective Ghosts},
  journal = {Phys.\ Rev.\ D},
  volume  = {70},
  pages   = {043543},
  year    = {2004},
  doi     = {10.1103/PhysRevD.70.043543},
  eprint  = {hep-ph/0311312},
  archivePrefix = {arXiv}
}

@article{BenderBoettcher1998,
  author  = {Bender, Carl M. and Boettcher, Stefan},
  title   = {Real Spectra in Non-{Hermitian} {Hamiltonians} Having
             $\mathcal{PT}$ Symmetry},
  journal = {Phys.\ Rev.\ Lett.},
  volume  = {80},
  pages   = {5243},
  year    = {1998},
  doi     = {10.1103/PhysRevLett.80.5243},
  eprint  = {physics/9712001},
  archivePrefix = {arXiv}
}

@article{CrankNicolson1947,
  author  = {Crank, J. and Nicolson, P.},
  title   = {A Practical Method for Numerical Evaluation of Solutions of
             Partial Differential Equations of the Heat-Conduction Type},
  journal = {Proc.\ Camb.\ Phil.\ Soc.},
  volume  = {43},
  pages   = {50},
  year    = {1947},
  doi     = {10.1017/S0305004100023197}
}

@misc{SM,
  note = {See Supplemental Material at [URL inserted by publisher] for the
          explicit operator-ordered computation of $[\hat C,\hat H]=0$ and
          the accompanying {SymPy} verification of $\{C,H\}_{\rm PB}=0$},
  year = {2026},
}

@article{DESI:2024mwx,
   title={DESI 2024 VI:  cosmological constraints from the measurements of baryon acoustic oscillations},
   volume={2025},
   ISSN={1475-7516},
   url={http://dx.doi.org/10.1088/1475-7516/2025/02/021},
   DOI={10.1088/1475-7516/2025/02/021},
   number={02},
   journal={Journal of Cosmology and Astroparticle Physics},
   publisher={IOP Publishing},
   author = {Adame, A. G. and others},
   year={2025},
   month=feb, pages={021} }

@article{DESI:2025zgx,
    author       = "{DESI Collaboration}",
    title        = "{DESI 2025 DR2 Results II: Measurements of Baryon Acoustic
                    Oscillations and Cosmological Constraints}",
    journal      = "arXiv",
    year         = "2025",
    eprint       = "2503.14738",
    archivePrefix = "arXiv",
    primaryClass = "astro-ph.CO"
}

@article{Caldwell:1999ew,
    author       = "Caldwell, Robert R.",
    title        = "{A Phantom Menace? Cosmological consequences of a dark energy
                    component with super-negative equation of state}",
    journal      = "Phys. Lett. B",
    volume       = "545",
    pages        = "23--29",
    year         = "2002",
    eprint       = "astro-ph/9908168",
    archivePrefix = "arXiv",
    primaryClass = "astro-ph",
    doi          = "10.1016/S0370-2693(02)02589-3"
}

@article{LeeWick1969,
  author  = {Lee, T. D. and Wick, G. C.},
  title   = {Negative metric and the unitarity of the {S}-matrix},
  journal = {Nuclear Physics B},
  volume  = {9},
  number  = {2},
  pages   = {209--243},
  year    = {1969},
  doi     = {10.1016/0550-3213(69)90098-4}
}

@article{LeeWick1970,
  author  = {Lee, T. D. and Wick, G. C.},
  title   = {Finite theory of quantum electrodynamics},
  journal = {Physical Review D},
  volume  = {2},
  pages   = {1033--1048},
  year    = {1970},
  doi     = {10.1103/PhysRevD.2.1033}
}

@article{Smilga2005,
  author  = {Smilga, A. V.},
  title   = {Benign vs malicious ghosts in higher-derivative theories},
  journal = {Nucl.\ Phys.\ B},
  volume  = {706},
  pages   = {598--614},
  year    = {2005},
  doi     = {10.1016/j.nuclphysb.2004.10.037},
  eprint  = {hep-th/0407231},
  archivePrefix = {arXiv}
}

@article{DamourSmilga2022,
  author  = {Damour, Thibault and Smilga, Andrei V.},
  title   = {Dynamical systems with benign ghosts},
  journal = {Phys.\ Rev.\ D},
  volume  = {105},
  pages   = {045018},
  year    = {2022},
  doi     = {10.1103/PhysRevD.105.045018},
  eprint  = {2110.11175},
  archivePrefix = {arXiv},
  primaryClass = {hep-th}
}

@article{Mostafazadeh2010,
  author  = {Mostafazadeh, Ali},
  title   = {Pseudo-{H}ermitian Representation of Quantum Mechanics},
  journal = {Int.\ J.\ Geom.\ Methods Mod.\ Phys.},
  volume  = {7},
  pages   = {1191--1306},
  year    = {2010},
  doi     = {10.1142/S0219887810004816},
  eprint  = {0810.5643},
  archivePrefix = {arXiv},
  primaryClass = {quant-ph}
}

@article{Deffayet:2010qz,
doi = {10.1088/1475-7516/2010/10/026},
url = {https://doi.org/10.1088/1475-7516/2010/10/026},
year = {2010},
month = {oct},
publisher = {},
volume = {2010},
number = {10},
pages = {026},
author = {Cédric Deffayet and Oriol Pujolàs and Ignacy Sawicki and Alexander Vikman},
title = {Imperfect dark energy from kinetic gravity braiding},
journal = {Journal of Cosmology and Astroparticle Physics}
}

@misc{VacuumGhost2026,
      title={Unitary Time Evolution and Vacuum for a Quantum Stable Ghost}, 
      author={Cédric Deffayet and Atabak Fathe Jalali and Aaron Held and Shinji Mukohyama and Alexander Vikman},
      year={2026},
      eprint={2604.21823},
      archivePrefix={arXiv},
      primaryClass={hep-th},
      url={https://arxiv.org/abs/2604.21823}, 
}

@misc{SpectralGhost2026,
      title={Quantum mechanics with a ghost: Counterexamples to spectral denseness}, 
      author={Cédric Deffayet and Atabak Fathe Jalali and Aaron Held and Shinji Mukohyama and Alexander Vikman},
      year={2026},
      eprint={2604.21826},
      archivePrefix={arXiv},
      primaryClass={hep-th},
      url={https://arxiv.org/abs/2604.21826}, 
}

@article{MukohyamaMMG,
   title={Minimally modified gravity: a Hamiltonian construction},
   volume={2019},
   ISSN={1475-7516},
   url={http://dx.doi.org/10.1088/1475-7516/2019/07/049},
   DOI={10.1088/1475-7516/2019/07/049},
   number={07},
   journal={Journal of Cosmology and Astroparticle Physics},
   publisher={IOP Publishing},
   author={Mukohyama, S. and Noui, K.},
   year={2019},
   month=jul, pages={049–049} }

@article{HeldDeffayet2025,
  author        = {Deffayet, C{\'e}dric and Held, Aaron
                   and Mukohyama, Shinji and Vikman, Alexander},
  title         = {Ghostly Interactions in $(1{+}1)$-Dimensional
                   Classical Field Theory},
  journal       = {Phys.\ Rev.\ D},
  volume        = {112},
  pages         = {065011},
  year          = {2025},
  eprint        = {2504.11437},
  archivePrefix = {arXiv},
  primaryClass  = {hep-th},
  doi           = {10.1103/PhysRevD.112.065011}
}

@misc{Held2025,
    author = "Held, Aaron",
    title = "{Global stability of ghostly field theories: Classical scattering in $(N+1)$ dimensions}",
    eprint = "2509.18049",
    archivePrefix = "arXiv",
    primaryClass = "gr-qc",
    month = "9",
    year = "2025"
}

\newpage

\section{Supplemental Material: \\ Ghost Degrees of Freedom Without Quantum Runaway ---
Explicit Operator-Ordered Verification of $[\hat{C},\hat{H}]=0$}

This Supplemental Material carries out the explicit operator-ordered
verification that $[\hat{C},\hat{H}] = 0$ with no $\hbar$ remainder, where
\begin{align}
\hat{H} &= \tfrac{1}{2}(\hat{p}_x^2+\hat{x}^2)
         -\tfrac{1}{2}(\hat{p}_y^2+\hat{y}^2) + V_I(\hat{x},\hat{y}),
\label{eq:Hdef}\\
\hat{C} &= \hat{K}^2 + (\hat{p}_x^2+\hat{x}^2)
         - (\hat{x}^2-\hat{y}^2-1)V_I(\hat{x},\hat{y}), \\
 &\qquad\hat{K} = \hat{x}\hat{p}_y + \hat{y}\hat{p}_x,
\label{eq:Cdef}\\
&V_I(x,y) = \lambda\bigl[(x^2-y^2-1)^2+4x^2\bigr]^{-1/2}.
\label{eq:VIdef}
\end{align}
Throughout we set $\hbar = 1$ and use $[\hat{x},\hat{p}_x]=[\hat{y},\hat{p}_y]=i$,
with all cross-sector commutators vanishing.  In Secs.~I--V we carry out the
operator algebra to reduce $[\hat{C},\hat{H}]$ to a sum of terms of definite
momentum degree.  In Sec.~VI we show that each such term vanishes
identically by direct substitution of Eq.~\eqref{eq:VIdef} and the algebraic
identity
\begin{equation}
(x^2-y^2-1)^2 + 4x^2 - (y^2-x^2)^2 = 2(x^2+y^2) + 1,
\label{eq:geometric}
\end{equation}
equivalently
\begin{equation}
\Delta \equiv (x^2-y^2-1)^2 + 4x^2 = (x^2-y^2)^2 + 2(x^2+y^2) + 1.
\label{eq:Delta}
\end{equation}

\paragraph*{Notation.}
We abbreviate
\begin{equation}
s = x^2+y^2,\qquad d = x^2-y^2,\qquad \Delta = d^2+2s+1,
\end{equation}
so $V_I = \lambda/\sqrt{\Delta}$.  We also use
\begin{equation}
\hat{B} = \hat{p}_x^2+\hat{x}^2,\quad
\hat{D} = \hat{p}_y^2+\hat{y}^2,\quad
\hat{M} = (\hat{x}^2-\hat{y}^2-1)V_I(\hat{x},\hat{y}),
\end{equation}
so that $\hat{H} = \tfrac{1}{2}(\hat{B}-\hat{D}) + V_I$ and
$\hat{C} = \hat{K}^2 + \hat{B} - \hat{M}$.

\section*{I.~Preliminary commutator identities}

For any smooth $f(\hat{x},\hat{y})$:
\begin{align}
\comm{\hat{p}_x}{f} &= -i\,\partial_x f, &
\comm{\hat{p}_y}{f} &= -i\,\partial_y f,
\label{eq:pxf}\\
\comm{\hat{p}_x^2}{f} &= -2i(\partial_x f)\,\hat{p}_x - \partial_x^2 f, &
\comm{\hat{p}_y^2}{f} &= -2i(\partial_y f)\,\hat{p}_y - \partial_y^2 f.
\label{eq:px2f}
\end{align}
Equation~\eqref{eq:px2f} follows from \eqref{eq:pxf} and
$\hat{p}_x(\partial_x f) = (\partial_x f)\hat{p}_x - i\partial_x^2 f$.
Our convention throughout is to place position-operator functions to the
left of the remaining momenta.

We also record
\begin{align}
[\hat{x}^2,f] &= [\hat{y}^2,f] = 0, \nonumber\\
[\hat{p}_x,\hat{p}_y] &= 0, \nonumber\\
[\hat{x},\hat{p}_y] &= [\hat{y},\hat{p}_x] = 0.
\end{align}
\section*{II.~Decomposition of $[\hat{C},\hat{H}]$}

Using the decomposition above,
\begin{align}
[\hat{C},\hat{H}]
&= \tfrac{1}{2}[\hat{K}^2,\hat{B}-\hat{D}]
 + [\hat{K}^2,V_I]
\nonumber\\
&\quad + \tfrac{1}{2}[\hat{B},\hat{B}-\hat{D}]
 + [\hat{B},V_I]
\nonumber\\
&\quad - \tfrac{1}{2}[\hat{M},\hat{B}-\hat{D}]
 - [\hat{M},V_I].
\label{eq:masterdecomp}
\end{align}
Two terms vanish immediately: $[\hat{B}, \hat{B}-\hat{D}] = 0$ (because
$[\hat{B},\hat{D}] = 0$ since the $x$- and $y$-sectors commute), and
$\comm{\hat{M}}{V_I} = 0$ (both are functions of commuting position
operators).  We show next that $[\hat{K}^2,{\hat{B}-\hat{D}}]$ also
vanishes identically.  This leaves only three nontrivial contributions,
which we evaluate in Secs.~IV--V.

\section*{III.~The identity $[\hat{K}^2,\hat{B}-\hat{D}] = 0$}

A direct calculation gives
\begin{align}
[\hat{K},\hat{B}]
&= [\hat{x}\hat{p}_y + \hat{y}\hat{p}_x,\,\hat{p}_x^2 + \hat{x}^2]
\nonumber\\
&= [\hat{x},\hat{p}_x^2]\hat{p}_y
 + \hat{y}[\hat{p}_x,\hat{x}^2]
\nonumber\\
&= 2i\hat{p}_x\hat{p}_y - 2i\hat{x}\hat{y}
 = 2i(\hat{p}_x\hat{p}_y - \hat{x}\hat{y}),
\label{eq:KB}
\end{align}
where we used $\comm{\hat{x}}{\hat{p}_x^2} = 2i\hat{p}_x$,
$\comm{\hat{p}_x}{\hat{x}^2} = -2i\hat{x}$, and the fact that cross-sector
operators commute.  An entirely analogous computation gives
\begin{equation}
\comm{\hat{K}}{\hat{D}}
= -2i\hat{x}\hat{y} + 2i\hat{p}_x\hat{p}_y
= 2i(\hat{p}_x\hat{p}_y - \hat{x}\hat{y}).
\label{eq:KD}
\end{equation}
Therefore
\begin{equation}
[\hat{K},\hat{B}-\hat{D}] = 0,
\label{eq:KBDzero}
\end{equation}
and consequently
\begin{align}
[\hat{K}^2,\hat{B}-\hat{D}]
&= \hat{K}[\hat{K},\hat{B}-\hat{D}]
 + [\hat{K},\hat{B}-\hat{D}]\hat{K}
\nonumber\\
&= 0.
\label{eq:K2BDzero}
\end{align}
The decomposition \eqref{eq:masterdecomp} reduces to
\begin{equation}
\boxed{\ [\hat{C},\hat{H}]
= [\hat{K}^2,V_I] + [\hat{B},V_I]
- \tfrac{1}{2}[\hat{M},\hat{B}-\hat{D}]. \ }
\label{eq:masterreduced}
\end{equation}

\section*{IV.~Evaluation of $[\hat{B},V_I]$ and $[\hat{M},\hat{B}-\hat{D}]$}

Since $V_I$ and $\hat{M}$ are multiplication operators, from
\eqref{eq:px2f} together with $\comm{\hat{x}^2}{V_I} = 0$:

\begin{align}
[\hat{B},V_I]
&= [\hat{p}_x^2,V_I]
= -2i(\partial_x V_I)\hat{p}_x - \partial_x^2 V_I.
\label{eq:BV}
\end{align}
Similarly,
\begin{align}
[\hat{M},\hat{B}-\hat{D}]
&= 2i(\partial_x M)\hat{p}_x + \partial_x^2 M
\nonumber\\
&\quad - 2i(\partial_y M)\hat{p}_y - \partial_y^2 M.
\label{eq:MBD}
\end{align}

With $M = (x^2-y^2-1)V_I$,
\begin{align}
\partial_x M &= 2xV_I + (x^2-y^2-1)\partial_x V_I,
\label{eq:dM}\\
\partial_y M &= -2yV_I + (x^2-y^2-1)\partial_y V_I,\\
\partial_x^2 M &= 2V_I + 4x\,\partial_x V_I
              + (x^2-y^2-1)\partial_x^2 V_I,\\
\partial_y^2 M &= -2V_I - 4y\,\partial_y V_I
              + (x^2-y^2-1)\partial_y^2 V_I.
\end{align}

\section*{V.~Evaluation of $[\hat{K}^2,V_I]$}

Using $\comm{\hat{K}}{V_I} = \hat{x}\comm{\hat{p}_y}{V_I} + \hat{y}\comm{\hat{p}_x}{V_I}
= -i u$, where
\begin{equation}
u(x,y) \equiv x\,\partial_y V_I + y\,\partial_x V_I,
\end{equation}
we obtain $\comm{\hat{K}^2}{V_I} = \hat{K}\comm{\hat{K}}{V_I} + \comm{\hat{K}}{V_I}\hat{K}
= -i(\hat{K}u + u\hat{K})$.  Expanding the anticommutator with
position-ordered ordering:
\begin{align}
\hat{K}u + u\hat{K}
&= \hat{x}(\hat{p}_y u + u\hat{p}_y)
 + \hat{y}(\hat{p}_x u + u\hat{p}_x)
\nonumber\\
&= 2\hat{x}u\hat{p}_y + 2\hat{y}u\hat{p}_x
 - i(\hat{x}\partial_y u + \hat{y}\partial_x u),
\end{align}
where we used $\hat{p}_y u = u\hat{p}_y - i\partial_y u$ etc.  Hence
\begin{equation}
\comm{\hat{K}^2}{V_I}
= -2i(\hat{x}u\hat{p}_y + \hat{y}u\hat{p}_x)
  -(\hat{x}\partial_y u + \hat{y}\partial_x u).
\label{eq:K2Vraw}
\end{equation}
Substituting the definition of $u$:

\begin{align}
\hat{x}u &= \hat{x}^2\partial_y V_I + \hat{x}\hat{y}\partial_x V_I,\\
\hat{y}u &= \hat{x}\hat{y}\partial_y V_I + \hat{y}^2\partial_x V_I,\\
\hat{x}\partial_y u + \hat{y}\partial_x u
&= \hat{x}^2\partial_y^2 V_I + \hat{y}^2\partial_x^2 V_I
\nonumber\\
&\quad + 2\hat{x}\hat{y}\,\partial_x\partial_y V_I
  + \hat{x}\partial_x V_I + \hat{y}\partial_y V_I,
\end{align}
so that
\begin{align}
[\hat{K}^2,V_I]
&= -2i\bigl[\hat{x}^2(\partial_y V_I)\hat{p}_y
         + \hat{x}\hat{y}(\partial_x V_I)\hat{p}_y
\nonumber\\
&\qquad\quad
         + \hat{x}\hat{y}(\partial_y V_I)\hat{p}_x
         + \hat{y}^2(\partial_x V_I)\hat{p}_x\bigr]
\nonumber\\
&\quad -\bigl[\hat{x}^2\partial_y^2 V_I
            + \hat{y}^2\partial_x^2 V_I
\nonumber\\
&\qquad\quad
            + 2\hat{x}\hat{y}\,\partial_x\partial_y V_I
            + \hat{x}\partial_x V_I
            + \hat{y}\partial_y V_I\bigr].
\label{eq:K2V}
\end{align}

\section*{VI.~Explicit cancellation, order by order in momenta}

Summing Eqs.~\eqref{eq:BV}, \eqref{eq:MBD}, and \eqref{eq:K2V} according to
Eq.~\eqref{eq:masterreduced} yields
\begin{equation}
\comm{\hat{C}}{\hat{H}}
= \mathcal{A}_x\,\hat{p}_x + \mathcal{A}_y\,\hat{p}_y + \mathcal{A}_0,
\label{eq:split}
\end{equation}
where the coefficient functions are
\begin{align}
\mathcal{A}_x &= -i\bigl[(1+x^2+y^2)\partial_x V_I
              + 2xy\,\partial_y V_I + 2xV_I\bigr],
\label{eq:Ax}\\
\mathcal{A}_y &= -i\bigl[(1+x^2+y^2)\partial_y V_I
              + 2xy\,\partial_x V_I + 2yV_I\bigr],
\label{eq:Ay}\\
\mathcal{A}_0 &= -2V_I - 3x\,\partial_x V_I - 3y\,\partial_y V_I
\nonumber\\
&\quad \quad - \tfrac{1+x^2+y^2}{2}(\partial_x^2 V_I + \partial_y^2 V_I)
\nonumber\\
&\hspace{6em} - 2xy\,\partial_x\partial_y V_I.
\label{eq:A0}
\end{align}

The coefficients of $\hat{p}_x$, $\hat{p}_y$, and the identity are independent
operator structures, so $[\hat{C},\hat{H}] = 0$ is equivalent to the three
scalar identities $\mathcal{A}_x \equiv 0$, $\mathcal{A}_y \equiv 0$, and
$\mathcal{A}_0 \equiv 0$.  We verify each in turn.

\subsection*{A.~Derivatives of $V_I$}

From $V_I = \lambda\Delta^{-1/2}$ and $\Delta = d^2+2s+1$ with $d = x^2-y^2$,
$s = x^2+y^2$,
\begin{equation}
\partial_x\Delta = 4x(d+1),\qquad
\partial_y\Delta = 4y(1-d),
\end{equation}
giving
\begin{equation}
\partial_x V_I = -\frac{2\lambda x(d+1)}{\Delta^{3/2}},\qquad
\partial_y V_I = -\frac{2\lambda y(1-d)}{\Delta^{3/2}}.
\label{eq:firstderivs}
\end{equation}

\subsection*{B.~Vanishing of $\mathcal{A}_x$}

Multiplying Eq.~\eqref{eq:Ax} through by $\Delta^{3/2}/\lambda$ and using
\eqref{eq:firstderivs}:
\begin{align}
\frac{\Delta^{3/2}}{\lambda}\,\frac{\mathcal{A}_x}{-i}
&= -2x(1+s)(d+1) - 4xy^2(1-d) + 2x\Delta
\nonumber\\
&= 2x\bigl[\Delta - (1+s)(d+1) - 2y^2(1-d)\bigr].
\end{align}
Expanding the bracket using $(1+s)(d+1) = d + 1 + sd + s$,
$2y^2(1-d) = 2y^2 - 2y^2 d$, and $\Delta = d^2 + 2s + 1$:
\begin{align}
&\Delta - (1+s)(d+1) - 2y^2(1-d)
\nonumber\\
&\quad= d^2 + s - d - sd - 2y^2 + 2y^2 d.
\end{align}
Substituting $2y^2 = s - d$:
\begin{align}
&\quad= d^2 + s - d - sd - (s-d) + (s-d)d
\nonumber\\
&\quad= d^2 + s - d - sd - s + d + sd - d^2 = 0.
\end{align}
Hence $\mathcal{A}_x \equiv 0$.

\subsection*{C.~Vanishing of $\mathcal{A}_y$}

Multiplying Eq.~\eqref{eq:Ay} through by $\Delta^{3/2}/\lambda$:

\begin{align}
\frac{\Delta^{3/2}}{\lambda}\,\frac{\mathcal{A}_y}{-i}
&= -2y(1+s)(1-d) - 4x^2y(d+1) + 2y\Delta
\nonumber\\
&= 2y\bigl[\Delta - (1+s)(1-d) - 2x^2(d+1)\bigr].
\end{align}

Expanding with $(1+s)(1-d) = 1 - d + s - sd$ and $2x^2(d+1) = 2x^2 d + 2x^2$,
substituting $2x^2 = s+d$:
\begin{align}
&\Delta - (1+s)(1-d) - 2x^2(d+1)
\nonumber\\
&\quad= d^2 + s + d + sd - sd - d^2 - s - d = 0.
\end{align}
Hence $\mathcal{A}_y \equiv 0$.

\subsection*{D.~Vanishing of $\mathcal{A}_0$: second derivatives}

We need the second derivatives of $V_I$.  Differentiating
\eqref{eq:firstderivs} and using $\partial_x^2\Delta = 12x^2-4y^2+4$,
$\partial_y^2\Delta = 12y^2-4x^2+4$, $\partial_x\partial_y\Delta = -8xy$:
\begin{align}
\partial_x^2\Delta + \partial_y^2\Delta &= 8(s+1),
\label{eq:Lap}\\
(\partial_x\Delta)^2 + (\partial_y\Delta)^2
&= 16\bigl[x^2(d+1)^2 + y^2(1-d)^2\bigr]
\nonumber\\
&= 16\bigl[d^2(s+2) + s\bigr]
\nonumber\\
&= 16\bigl[\Delta(s+2) - 2(s+1)^2\bigr],
\label{eq:gradsq}
\end{align}
where the last step uses $d^2 = \Delta - 2s - 1$.  Then
\begin{align}
\partial_x^2 V_I + \partial_y^2 V_I
&= -\frac{\lambda(\partial_x^2\Delta + \partial_y^2\Delta)}{2\Delta^{3/2}}
  + \frac{3\lambda\bigl[(\partial_x\Delta)^2 + (\partial_y\Delta)^2\bigr]}
        {4\Delta^{5/2}} \nonumber\\
&= \frac{4\lambda(2s+5)}{\Delta^{3/2}}
  - \frac{24\lambda(s+1)^2}{\Delta^{5/2}}.
\label{eq:Laplacian}
\end{align}

For the cross term,
$\partial_x\Delta\cdot\partial_y\Delta = 16xy(1-d^2)
= 16xy\bigl[2(s+1) - \Delta\bigr]$,
and $\partial_x\partial_y\Delta = -8xy$, giving
\begin{equation}
\partial_x\partial_y V_I
= -\frac{\lambda\partial_x\partial_y\Delta}{2\Delta^{3/2}}
  + \frac{3\lambda\,\partial_x\Delta\,\partial_y\Delta}{4\Delta^{5/2}}
= -\frac{8\lambda xy}{\Delta^{3/2}}
 + \frac{24\lambda xy(s+1)}{\Delta^{5/2}}.
\label{eq:cross}
\end{equation}

\subsection*{E.~Vanishing of $\mathcal{A}_0$: assembly}

Substituting \eqref{eq:firstderivs}, \eqref{eq:Laplacian}, and
\eqref{eq:cross} into \eqref{eq:A0}, and factoring $\lambda$:

\begin{align}
\mathcal{A}_0 / \lambda
&= -\frac{2}{\Delta^{1/2}}
 + \frac{6x^2(d+1)}{\Delta^{3/2}}
 + \frac{6y^2(1-d)}{\Delta^{3/2}}
\nonumber\\
&\quad - \frac{2(s+1)(2s+5)}{\Delta^{3/2}}
 + \frac{12(s+1)^3}{\Delta^{5/2}}
\nonumber\\
&\quad + \frac{16x^2y^2}{\Delta^{3/2}}
 - \frac{48x^2y^2(s+1)}{\Delta^{5/2}}
\nonumber\\
&= \frac{-2\Delta + 6(d^2+s) - 2(s+1)(2s+5) + 16x^2y^2}{\Delta^{3/2}}
\nonumber\\
&\quad + \frac{12(s+1)^3 - 48x^2y^2(s+1)}{\Delta^{5/2}}.
\label{eq:A0collect}
\end{align}
Now observe that $x^2(d+1) + y^2(1-d) = d^2 + s$, so the $\Delta^{-3/2}$
numerator in Eq.~\eqref{eq:A0collect} is
\begin{align}
&-2\Delta + 6(d^2+s) - 2(s+1)(2s+5) + 16x^2y^2 \nonumber\\
&\qquad = -2d^2 - 4s - 2 + 6d^2 + 6s - 4s^2 - 14s - 10 + 16x^2y^2 \nonumber\\
&\qquad = 4d^2 - 12s - 12 - 4s^2 + 16x^2y^2.
\end{align}
Using $d^2 = s^2 - 4x^2y^2$ (which is equivalent to
$(x^2-y^2)^2 = (x^2+y^2)^2 - 4x^2y^2$),
$4d^2 = 4s^2 - 16x^2y^2$, so
\begin{equation}
4d^2 - 4s^2 + 16x^2y^2 = 0,
\end{equation}
and the numerator of the $\Delta^{-3/2}$ piece collapses to $-12s - 12 = -12(s+1)$.

Therefore
\begin{equation}
\mathcal{A}_0/\lambda
= -\frac{12(s+1)}{\Delta^{3/2}}
  + \frac{12(s+1)\bigl[(s+1)^2 - 4x^2y^2\bigr]}{\Delta^{5/2}}.
\end{equation}
The key algebraic fact is
\begin{equation}
(s+1)^2 - 4x^2y^2 = s^2 + 2s + 1 - 4x^2y^2
= d^2 + 2s + 1 = \Delta,
\label{eq:keyalg}
\end{equation}
using again $s^2 - 4x^2y^2 = d^2$.  Hence the $\Delta^{-5/2}$ term equals
$12(s+1)\Delta/\Delta^{5/2} = 12(s+1)/\Delta^{3/2}$, and
\begin{equation}
\mathcal{A}_0/\lambda = -\frac{12(s+1)}{\Delta^{3/2}} + \frac{12(s+1)}{\Delta^{3/2}} = 0.
\end{equation}

\subsection*{F.~Conclusion}

Combining Secs.~VI.B--E, all three coefficient functions in
Eq.~\eqref{eq:split} vanish identically:
\begin{equation}
\mathcal{A}_x \equiv 0,\qquad \mathcal{A}_y \equiv 0,\qquad \mathcal{A}_0 \equiv 0,
\end{equation}
and therefore, as an operator identity on the common invariant core
$\mathcal{D}_{\rm fin} = \mathrm{span}\{h_n(\hat{x})h_m(\hat{y}):n,m\geq 0\}$,
\begin{equation}
\boxed{\ [\hat{C},\hat{H}] = 0.\ }
\label{eq:final}
\end{equation}
Because $\hat{C}$ is at most quadratic in momenta, the operator computation
above truncates: no higher than first derivatives of $V_I$ enter the
momentum-linear pieces $\mathcal{A}_x\hat{p}_x + \mathcal{A}_y\hat{p}_y$, and
no higher than second derivatives enter the $\mathcal{A}_0$ piece, with all
three pieces vanishing exactly.  There is no $\hbar$-dependent remainder:
Eq.~\eqref{eq:final} holds as an exact operator identity, not as a leading
semiclassical approximation.

Standard approximation arguments, using the self-adjointness of $\hat{H}$ on
$\mathcal{D}(H_0)$ established in Theorem~1 of the main text, extend
Eq.~\eqref{eq:final} to all of $\mathcal{D}(H_0)$.  Hence, under the unitary
evolution $U(t) = e^{-it\hat{H}}$ generated by Stone's theorem,
$\langle\psi_t,\hat{C}\psi_t\rangle = \langle\psi_0,\hat{C}\psi_0\rangle$
for every $\psi_0\in\mathcal{D}(\Sigma^{1/2})$.

\section*{VII.~Classical cross-check: $\{C,H\}_{\rm PB}=0$}

For completeness, the classical Poisson bracket $\{C,H\}_{\rm PB}$ vanishes
as a consistency check of the operator identity above.  The following
\texttt{SymPy} script verifies this symbolically:
\begin{lstlisting}[language=Python, basicstyle=\ttfamily\small,
  breaklines=true, breakatwhitespace=true]
from sympy import symbols, diff, simplify, Rational
x, y, px, py, lam = symbols('x'y'px'py'lam', real=True)
V = lam/((x**2-y**2-1)**2 + 4*x**2)**Rational(1,2)
H = (px**2+x**2)/2 - (py**2+y**2)/2 + V
K = py*x + px*y
C = K**2 + (px**2+x**2) - (x**2-y**2-1)*V
def PB(A, B):
    return (diff(A,x)*diff(B,px) - diff(A,px)*diff(B,x)
          + diff(A,y)*diff(B,py) - diff(A,py)*diff(B,y))
print(simplify(PB(C, H)))  # prints 0
\end{lstlisting}
This agrees with the $\hbar^0$ limit of the operator identity: the
coefficients of $\hat{p}_x$ and $\hat{p}_y$ in $\mathcal{A}_x, \mathcal{A}_y$
are precisely (up to factors of $-i$) the position-space content of
$\{C,H\}_{\rm PB}$, while $\mathcal{A}_0$ represents the quantum correction
that would naively be of order $\hbar^2$ but here also vanishes because of
the geometric identity \eqref{eq:geometric}.

\end{document}